\documentclass[12pt]{article}
\usepackage{graphicx}
\input{epsf}
\begin{document}

\title{ 
Radiative corrections to semi-inclusive deep inelastic scattering
induced by lepton and photon pair electroproduction
}
\author{ A. Ilyichev\thanks{E-mail:ily@hep.by}\\
{\small National Scientific
and Educational Center of Particle and High Energy Physics,}\\ 
{\small 
 Belarusian State University,
220088  Minsk,  Belarus}\\
M. Osipenko\\
{\small Istituto Nazionale di Fisica Nucleare, Sezione di Genova,
	16146 Genoa, Italy
}}
\date{}
\maketitle
\begin{abstract}
The contribution of lepton and photon pair production in $ep$-scattering 
to the cross section of semi-inclusive deep inelastic scattering $ep \to e^\prime p^\prime X$
has been calculated.
The numerical results showed a large contribution of this processes at $\phi_h=180^o$,
in a good agreement with preliminary experimental data.
\end{abstract}

\section{Introduction}
Semi-inclusive deep inelastic scattering (SIDIS) allows to investigate not only 
parton distributions but also shed light on parton hadronization process.
During the data analysis of this reaction it is necessary to account for the QED radiative corrections
spoiling experimental observables. 

The lowest order radiative corrections to SIDIS were first calculated
for the scattering of the polarized particles and three-fold cross section in~\cite{sirad}.
These calculations allowed to develop the code SIRAD, which represented
a modification of DIS program POLRAD~\cite{polrad}.
Then in~\cite{haprad} the calculation of radiative corrections was generalized
for unpolarized five-fold cross section and code HAPRAD was developed.
In all these calculations Bardin-Shumeiko covariant approach~\cite{BSh}
was used for cancellation of infrared divergences. 
At last in~\cite{haprad20} the contribution of exclusive radiative tail was studied
and a new version of HAPRAD (named {\bf 2.0}) was written\footnote{All of these code can be found
at http://www.hep.by/rc}.

In the present work we consider another background process for a specific semi-inclusive reaction,
in which the detected hadron is identical to the initial one.
Preliminary experimental data from JLab~\cite{JLab} showed that in the kinematic region
near $\phi_h=180^o$ the measured cross section exceeded the calculated one-photon exchange (Born) contribution
and the excess could not be described by the radiative corrections calculated before.  
It was noticed that the main contribution in this region came from the exclusive
lepton pair electroproduction. In spite of its higher QED order (suppressed by $\alpha^2$)
this process is enhanced by the square of proton electromagnetic form-factors at low $Q^2$.
Therefore, in some specific kinematics the contribution of this process compete
with SIDIS reaction cross section.

The matrix elements of the lepton pair electroproduction and the phase space of this process
are considered in the next section. The contribution from two photon emission are also considered there
as the process of the same order. Numerical result and conclusion presented in the next sections.

\begin{figure}
\vspace*{-23mm}
\hspace*{43mm}
\includegraphics[width=5cm,height=5cm]{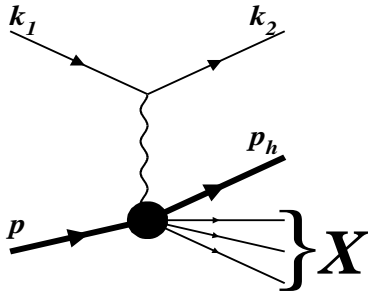}
\caption{
Semi-inclusive reaction.}
\label{fig0}
\end{figure}

\section{Method of calculation}
\label{mc}
Semi-inclusive reaction (see Fig.~{\ref{fig0}})
\begin{eqnarray}
e^-(k_1)+p(p)\rightarrow  e^-(k_2)+p(p_h)+X
\end{eqnarray}
can be describe by five variables:
\begin{eqnarray}
Q^2=-q^2=-(k_1-k_2)^2,\; x=\frac{Q^2}{2p q},\; x_p=1-\frac{p_hq}{pq},\;t=(p-p_h)^2,\;
\phi _h,
\end{eqnarray}
where $\phi _h$ is the angle between $({\bf k}_1,{\bf k}_2)$ and $({\bf k}_1-{\bf k}_2,{\bf p}_h)$
planes. 

\begin{figure}
\includegraphics[width=3cm,height=3cm]{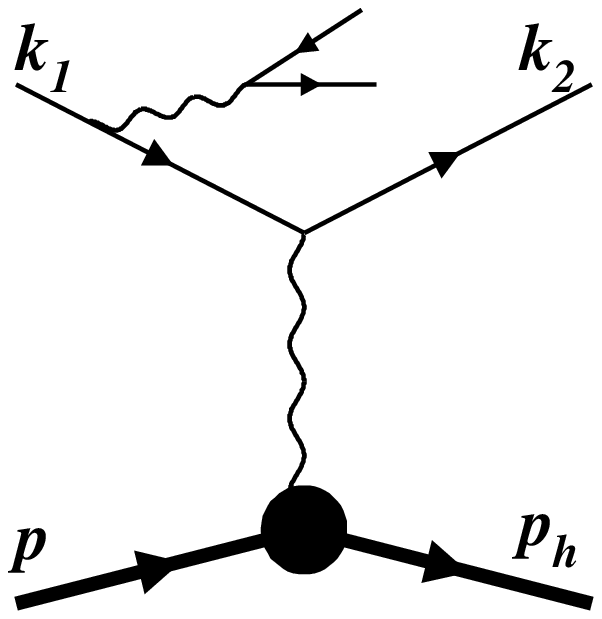}
\hspace*{-3.5mm}
\includegraphics[width=3cm,height=3cm]{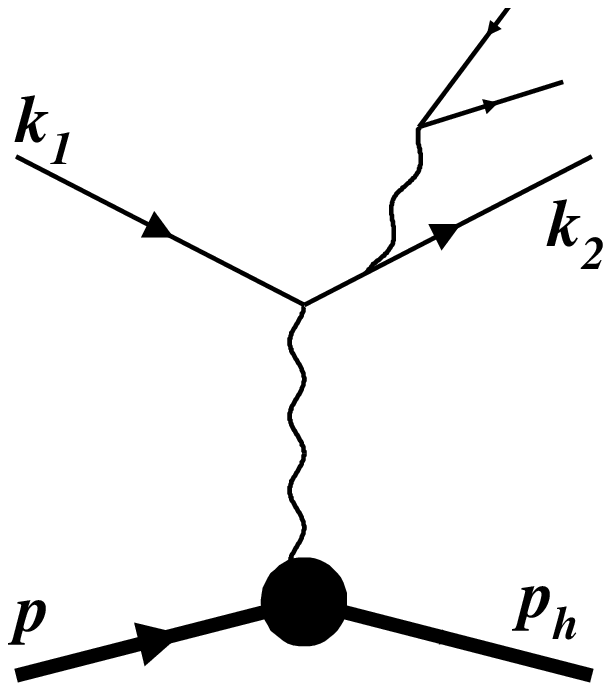}
\hspace*{-3.5mm}
\includegraphics[width=3cm,height=3cm]{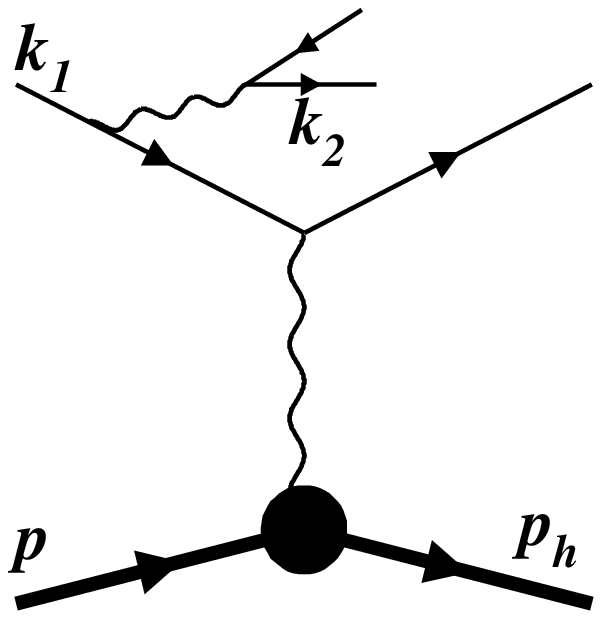}
\hspace*{-3.5mm}
\includegraphics[width=3cm,height=3cm]{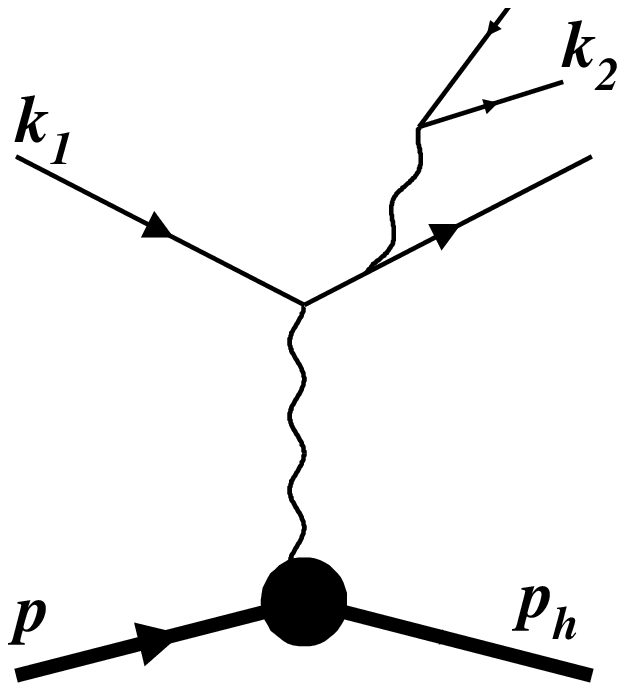}
\hspace*{-3.5mm}
\includegraphics[width=3cm,height=3cm]{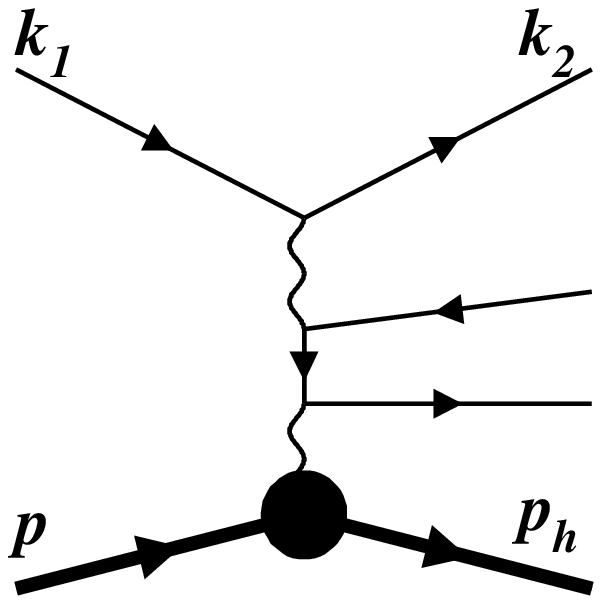}
\hspace*{2mm}
\includegraphics[width=3cm,height=3cm]{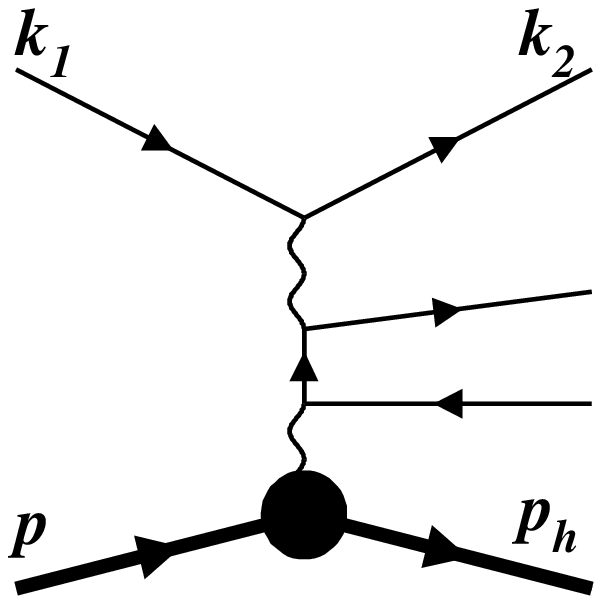}
\hspace*{2mm}
\includegraphics[width=3cm,height=3cm]{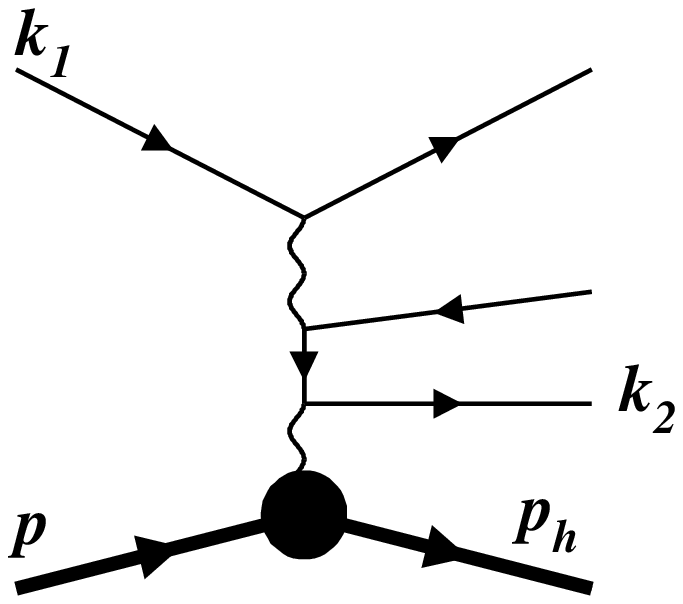}
\hspace*{2mm}
\includegraphics[width=3cm,height=3cm]{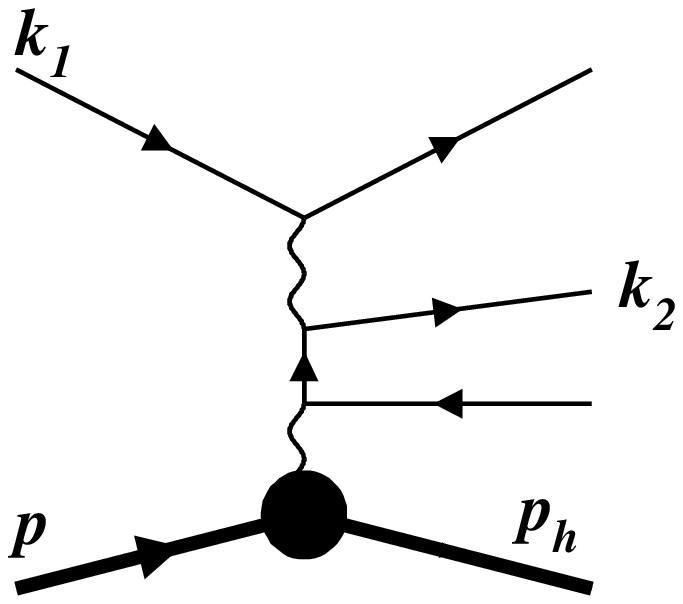}
\put(-298,110){\mbox{a)}}
\put(-103,110){\mbox{b)}}
\put(-305,-10){\mbox{c)}}
\put(-110,-10){\mbox{d)}}
\caption{The four pairs of gauge invariant graphs for the 
 lowest order of the lepton pair electroproduction
in $ep$-scattering.}
\label{fig1}
\end{figure}

The cross-section of electroproduction of the lepton pair in the $ep$-scattering
\begin{eqnarray}
e^-(k_1)+p(p)\rightarrow  e^-(k_2)+p(p_h)+e^-(l_-)+e^-(l_+)
\end{eqnarray}
at the lowest order reads: 
\begin{eqnarray}
d\sigma =\frac 1{2S} {\mathcal M}^{\dagger }{\mathcal M} d\Gamma ,
\label{cs}
\end{eqnarray}
with $S=2pk_1$.

We consider now the semi-inclusive process in which detected hadron is identical to the initial one.
In order to use SIDIS variables the phase space of this process can be transformed in the following way:
\begin{eqnarray}
d\Gamma =\frac 1{(2\pi)^8}
\frac {d^3k_2}{2k_{20}}
\frac {d^3p_h}{2p_{h0}}
d\Gamma _{2l}=
\frac {S_x^3dQ^2dxdx_pdtd\phi _h}{2^9\pi ^7	Q^2 S
\sqrt{S^2_x+4M^2Q^2}}
d\Gamma _{2l},
\label{dg}
\end{eqnarray}
where $S_x=2 pq$ and the phase space of the lepton pair has a form:
\begin{eqnarray}
d\Gamma _{2l}=
\frac {d^3l_+}{2l_{+0}}
\frac {d^3l_-}{2l_{-0}}\delta ^4(k_1+p-k_2-p_h-l_--l_+)=\frac 18 d\Omega _R,
\end{eqnarray}
with $d\Omega _R $ being the solid angle of lepton pair defined in its center mass system
$({\bf l}_++{\bf l}_-=0)$.

The matrix element can be written as a sum of the four pairs of gauge invariant
diagrams
${\mathcal M}={\mathcal M}_a+{\mathcal M}_b+{\mathcal M}_c+{\mathcal M}_d$
presented in Fig.~\ref{fig1}. The explicit expressions for each term read:
\begin{eqnarray}
{\mathcal M}_a&=&\frac{4\pi \alpha}{k^2q_h^2}{\bar u}(k_2)\biggl[
\gamma_\mu \frac{2k_{1\alpha}-{\hat k}\gamma_\alpha}{k^2-2k_1k}
+\frac{2k_{2\alpha}+{\gamma_\alpha\hat k}}{k^2+2k_2k}\gamma_\mu \biggr]u(k_1)
\nonumber \\&&\times
{\bar u}(l_-)\gamma_\alpha u(-l_+)J^h_\mu,
\nonumber \\
{\mathcal M}_c&=&\frac{4\pi \alpha}{q^2q_h^2}{\bar u}(l_-)\biggl[
\gamma_\mu \frac{{\hat q}\gamma_\alpha-2l_{+\alpha}}{q^2-2l_+q}
+\frac{2l_{-\alpha}+{\gamma_\alpha\hat q}}{q^2-2l_-q}\gamma_\mu \biggr]u(-l_+)
\nonumber \\&&\times
{\bar u}(k_2)\gamma_\alpha u(k_1)J^h_\mu,
\nonumber \\
{\mathcal M}_{b,d}&=& -{\mathcal M}_{a,c}(k_2\leftrightarrow  l_-),
\label{mll}
\end{eqnarray}
where $k=l_++l_-$, $q_h=q-k$,
\begin{eqnarray}
J^h_\mu={\bar u}(p_h)\biggl[ \gamma_\mu F_1+i\sigma _{\mu \nu }\frac{p_h^\nu-p^\nu}{2M} F_2 \biggr ] u(p)
\end{eqnarray}
and $F_{1,2}$ are the proton form factors.
\begin{figure}
\hspace*{15mm}
\includegraphics[width=3cm,height=3cm]{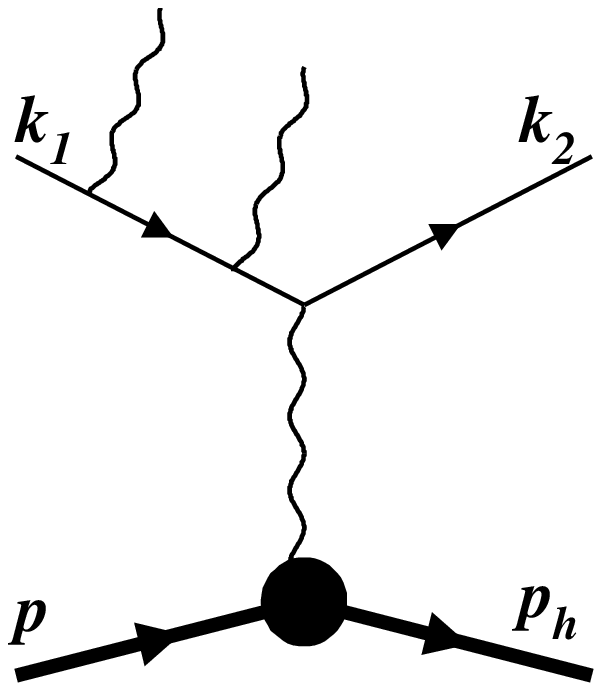}
\hspace*{-3mm}
\includegraphics[width=3cm,height=3cm]{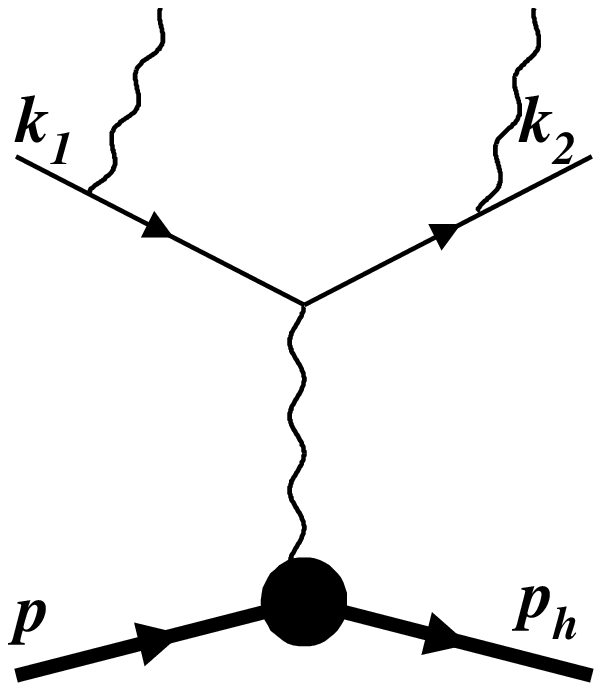}
\hspace*{-3mm}
\includegraphics[width=3cm,height=3cm]{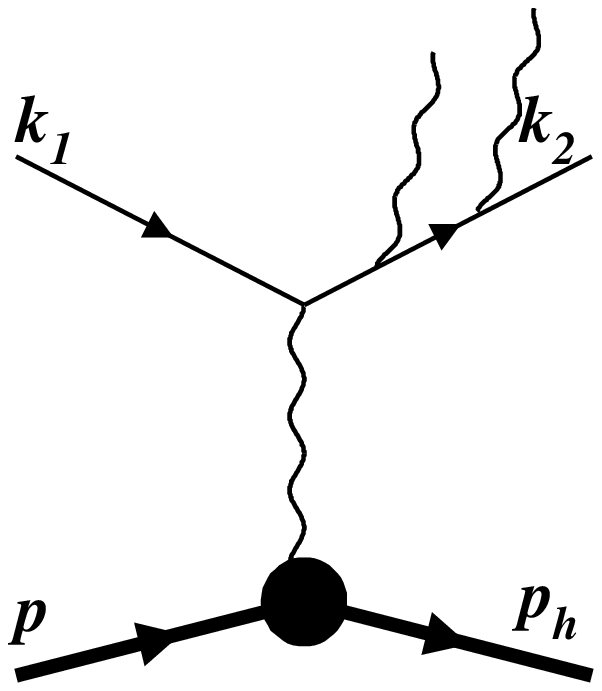}
\caption{The graphs for the 
two photon electroproduction
in $ep$-scattering.}
\label{fig2}
\end{figure}

Additional contribution of the same order comes from two photon emission which is shown in Fig.~\ref{fig2}:
\begin{eqnarray}
e^-(k_1)+p(p)\rightarrow  e^-(k_2)+p(p_h)+\gamma(\kappa _1,\varepsilon_1)+\gamma(\kappa _2,\varepsilon_2),
\end{eqnarray}
where $\varepsilon_{1,2}$ are photon polarization vectors.

The matrix element of this process reads:
\begin{eqnarray}
{\mathcal M}_{2\gamma }&=&\frac{4\pi \alpha}{q_h^2}{\bar u}(k_2)\biggl[
-\frac {(2k_{2\mu }+\gamma_\mu{\hat q}_h){\hat \varepsilon_2}(2k_1\varepsilon_1
-{\hat \kappa}_1{\hat \varepsilon}_1)}{2k_1\kappa_1(k^2-2k\kappa_1)} 
\nonumber \\&&
-\frac {(2k_2\varepsilon_2+{\hat \varepsilon_2}{\hat \kappa}_2)\gamma_\mu(2k_1\varepsilon_1
-{\hat \kappa}_1{\hat \varepsilon}_1)}{4k_1\kappa_1\;k_2\kappa_2} 
\nonumber \\&&
+\frac {(2k_2\varepsilon_2+{\hat \varepsilon_2}{\hat \kappa}_2){\hat \varepsilon}_1
\gamma_\mu(2k_{1\mu }-{\hat q}_h\gamma_\mu)}{2(k^2+2kk_2)k_2\kappa_2} 
\nonumber \\&&
+(\kappa _1,\varepsilon_1 \leftrightarrow \kappa _1,\varepsilon_1 )
\biggr]u(k_1),
\end{eqnarray}
with $k=\kappa _1+\kappa _2$, while the cross section and phase space looks like 
(\ref{dg}) and (\ref{cs}), respectively.

\section{Numerical results}
\begin{figure}
\includegraphics[bb=2cm 6cm 22cm 23cm,width=6cm,height=6cm]{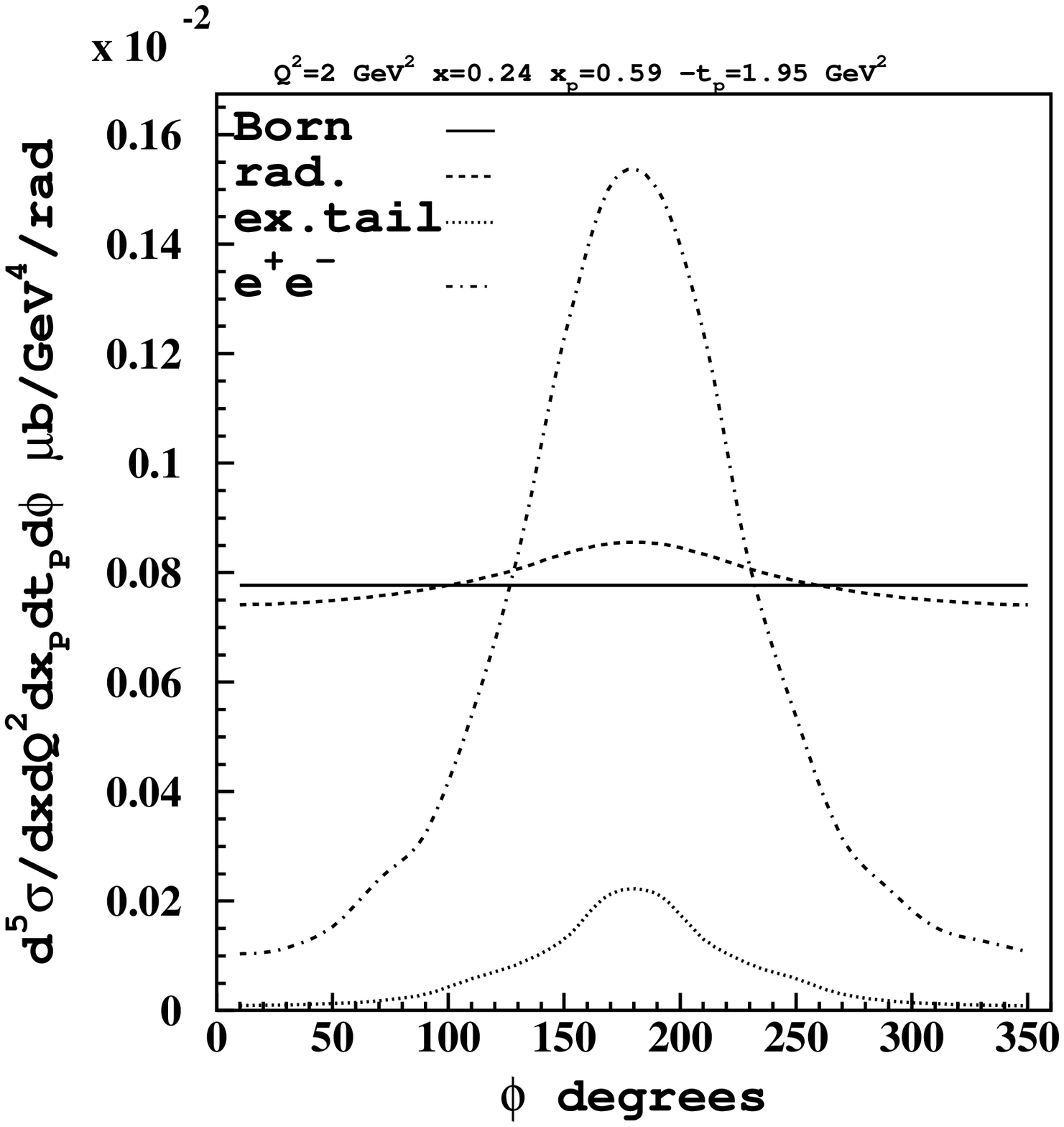}
\hspace*{3mm}
\includegraphics[bb=2cm 6cm 22cm 23cm,width=6cm,height=6cm]{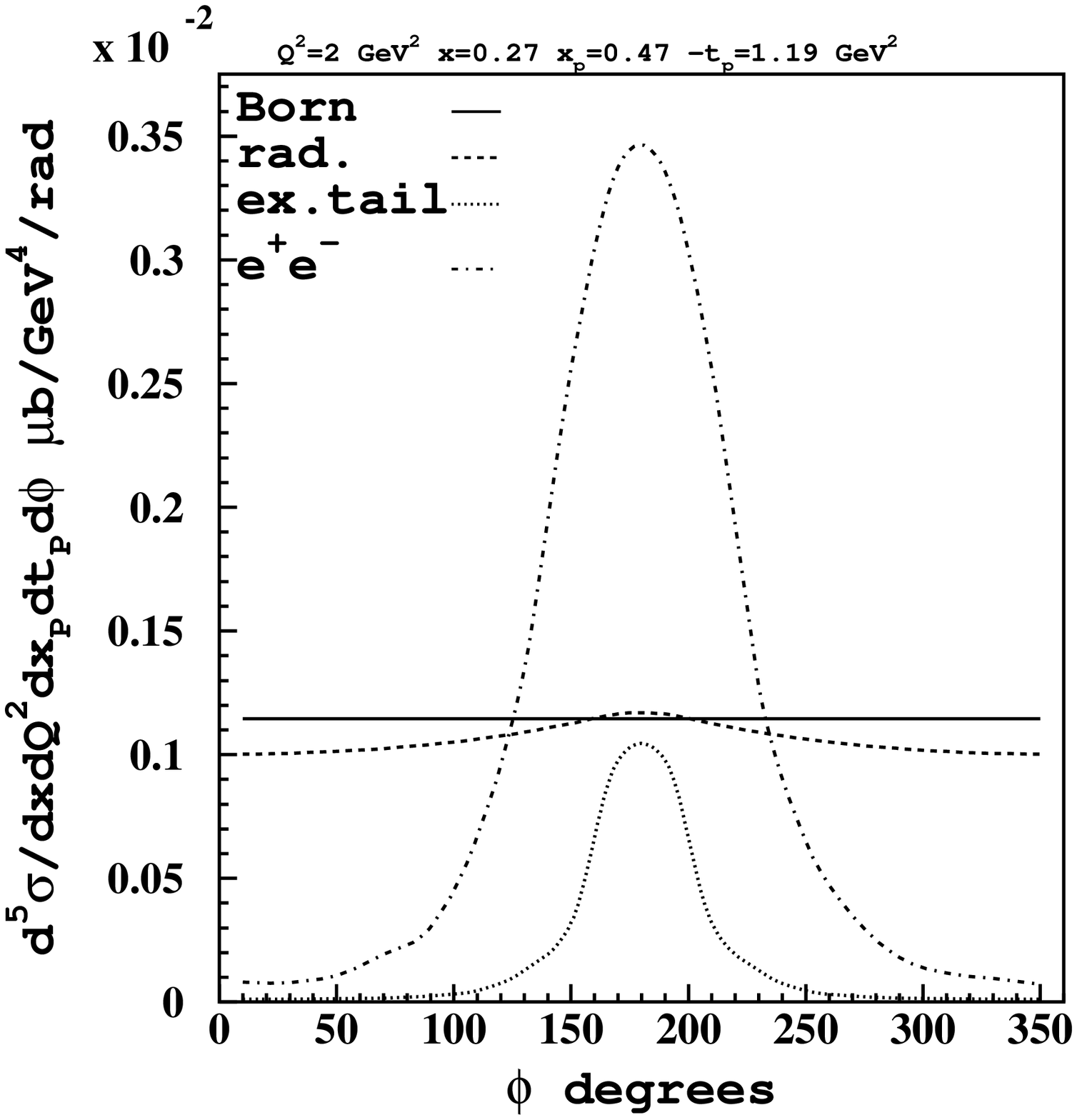}
\caption{$\phi _h$-dependence of the different contributions to the semi-inclusive 
process $ep\to e^\prime p^\prime X$. }
\label{fig3}
\end{figure}
The $\phi_h$-dependence of the different contributions to the semi-inclusive
reaction $e+p \to e^\prime + p^\prime + X$ is shown in Fig.~\ref{fig3}.
In the region near $\phi_h=180^o$ a narrow peak is observed.
Such behavior cannot be described by the Born contribution far as it
has $\phi_h$-dependence inconsistent with the usual unpolarized SIDIS distribution
$A+B\cos{\phi_h}+C\cos{2\phi_h}$. Radiative corrections calculated in
\cite{haprad} as well as exclusive radiative tail \cite{haprad20} ({\bf rad} and {\bf ex. tail} 
lines respectively) cannot improve this situation.   	
The dominant contribution in this region comes from the exclusive electroproduction of lepton pairs
in $ep$-scattering discussed above. Instead the two photon emission, shown in Fig.~\ref{fig2},
is almost negligible despite being of the same $\alpha$-order.
Moreover, detailed numerical estimates showed that most of correction strength
came from the square of matrix element ${\mathcal M}_b$ (see Fig.~\ref{fig1} (b) and equation (\ref{mll})).
These calculations are in good agreement with preliminary experimental data from~\cite{JLab}.

\section{Conclusion}
Radiative corrections to semi-inclusive electroproduction of proton induced by
the exclusive production of lepton and photon pairs have been calculated for the first time.
Numerical results showed important contribution (mainly from diagrams Fig.~\ref{fig1} (b))
in the region of $\phi _h=180^o$,
in good agreement with preliminary experimental data.
The presented approach is rather general and can be extended on analysis of
$A^{\prime} $-production~\cite{apr}.

\section*{Acknowledgments}
The one of authors (A.I.) would like to thanks INFN staff for generous hospitality during his visits.


\begin{thebibliography}{99}
\bibitem{sirad}
A.~V.~Soroko and N.~M.~Shumeiko,
  Sov.\ J.\ Nucl.\ Phys.\  {\bf 53}, 628 (1991)
  [Yad.\ Fiz.\  {\bf 53}, 1015 (1991)].
\bibitem{polrad}
I.~Akushevich, A.~Ilyichev, N.~Shumeiko, A.~Soroko and A.~Tolkachev,
  Comput.\ Phys.\ Commun.\  {\bf 104}, 201 (1997)
\bibitem{haprad}
I.~Akushevich, N.~Shumeiko and A.~Soroko,
  Eur.\ Phys.\ J.\ C {\bf 10}, 681 (1999)
\bibitem{BSh}
 D.~Y.~Bardin and N.~M.~Shumeiko,
  Nucl.\ Phys.\ B {\bf 127}, 242 (1977).
\bibitem{haprad20} 
I.~Akushevich, A.~Ilyichev and M.~Osipenko,
  Phys.\ Lett.\ B {\bf 672}, 35 (2009)
\bibitem{JLab} ``Study of Fracture Functions in Polarized and Unpolarized Semi-inclusive Electron Scattering off the Proton``,
M.~Osipenko et al., proposal for CLAS approved analysis (2005).
\bibitem{apr} 
J.D. Bjorken, R. Essig, P. Schuster, N. Toro, Phys. Rev. {\bf D80}, 075018 (2009). 
\end{thebibliography}
\end{document}